\documentclass[useAMS,usegraphicx,usenatbib]{mn2e}
\newcommand{\mc}{\multicolumn}
\title[Determination of age, luminosity and mass functions of young stars]
{On the determination of age, and mass functions of stars in 
young open star clusters from the analysis of their luminosity
functions}
\author[A.E.Piskunov et al.]
       {A.E.~Piskunov,$^1$\thanks{E-mail:
        piskunov@inasan.rssi.ru;\mbox{abelikov@ari.uni-heidelberg.de} 
        nkhar@mao.kiev.ua; sagar@upso.ernet.in; purni@iiap.res.in}
        A.N.~Belikov,$^1$
        N.V.~Kharchenko,$^2$
        R.~Sagar,$^{3,4}$
        A.~Subramaniam$^4$\\
        $^1$Institute of Astronomy of the Russian Acad. Sci.,
	          48 Pyatnitskaya Str., Moscow 119017, Russia\\
        $^2$Main Astronomical Observatory, 27 Akademika Zabolotnogo Str.,
             03680  Kiev, Ukraine\\ 
        $^3$State Observatory, Manora Peak, Naini Tal -- 263129, India\\
        $^4$Indian Institute of Astrophysics, Bangalore 560034, India}
\date{Accepted 2003 ... ...
       Received 2003 ... ..;
       in original form 2003 ... ..}

\pagerange{\pageref{firstpage}--\pageref{lastpage}}
\pubyear{2003}

\begin{document}

\maketitle

\label{firstpage}

\begin{abstract}   Based on the CCD observations of remote young open
clusters NGC~2383, NGC~2384, NGC~4103, NGC~4755, NGC~7510 and
Hogg~15,  we construct their observed luminosity functions (LFs). The
observed LFs are corrected for field star contamination determined
with the help of  galactic star count model. In the case of Hogg~15
and NGC~2383 we also  considered the additional contamination from
neighboring clusters NGC~4609 and NGC~2384 respectively. These
corrections provided the realistic pattern of cluster LF in the
vicinity of the MS turn on point and at fainter magnitudes, revealed
the so called H-feature arising  due to transition of the Pre-MS
phase to MS, which is dependent on the  cluster age. The theoretical
LFs were constructed representing a cluster  population model with
continuous star formation for a short time scale and  a power law
Initial Mass Function (IMF) and these were fitted to the  observed
LF. As a result we are able to determine for each cluster a set  of
parameters, describing cluster population (the age, duration of star 
formation, IMF slope, and percentage of field star contamination). It
was found that in spite of the non-monotonic behavior of observed
LFs, cluster IMFs can be  described as the power law functions with
slopes similar to Salpeter's  value. The present Main Sequence turn
on cluster ages are several times  lower than those derived from the
fitting of theoretical isochrones to  turn off region of the upper
Main Sequences.
\end{abstract}


\section{Introduction}

Young clusters are natural laboratories for the study of various
issues related to star formation processes, e.g., the initial mass
function (IMF), duration of star formation etc. In recent years, some
remote young  clusters have been studied using $BVRI$ CCD 
observations (e.g., see \citealt{sg91}, \citealt{phj},
\citealt{sc95},  \citealt{sc97}, \citealt{sg98}, \citealt{smb},
\citealt{ss99},  \citealt{san}, \citealt{pand}, \citealt{pia02}).
These observations were used for the  construction of color-color
diagrams, color-magnitude diagrams (CMDs) and  luminosity functions
(LFs). These diagrams were used for the determination  of cluster
reddening, distance and age while the LFs were transformed to stellar
mass spectrum, which can be identified with the IMF for  young (age
$\le$ 100 Myr) star clusters. In these conventional studies of young
star clusters, cluster age, LFs and MFs could not be determined 
accurately due to reasons discussed below.

Firstly, for reliable separation of cluster members from the
foreground and  background field stars present in the direction of 
distant ($d>1$ kpc) clusters, accurate kinematic (proper motion
and/or radial velocity) measurements  are essential.  Unfortunately,
even present day most accurate  Hipparcos proper motion data
(available for brightest $V<13$ mag stars only) are not sufficient
for this purpose. In such circumstances the field star contamination 
is removed statistically by observing comparison areas (blank fields)
adjacent to the cluster field.  It should be noted, however, that in
most cases of the conventional general-purpose CCD photometry, which
are not specially intended to e.g. the IMF study, such regions are
not observed, and hence even statistical means of the data refining
could not be used. This leads to a loss of information suitable
otherwise for the LF construction.

The other problem in the CMDs of young star cluster is the precise
determination of the Main Sequence (MS) turn off point, which is used
for age determination, since for young clusters due to their small
age, the upper MS is practically unevolved. Consequently, it becomes
difficult to evaluate the accurate cluster age, and instead an upper
estimate of the cluster age is provided. The difficulty is further 
increased by effects like stellar rotation and binarity, and by the 
morphology of the upper MS, which is practically vertical for early
type stars in the colours $(B-V)$, $(V-R)$, or $(V-I)$. 

These problems can be avoided if the cluster age is estimated using
the location of the MS turn on point, the place where Pre-MS stars
join the MS in the cluster CMD. For a young coeval cluster this point
resides about 7 mag fainter the brightest termination point of the
MS. As a result of the deep CCD observations of the clusters, this
region  can be seen in the CMD, and hence is available for the
analysis. Many of the above effects (except that of unresolved
binaries) are much weaker in the turn-on point region, which makes
the turn-on dating method to be more attractive than the turn off
one. There are, however, practical problems in locating the turn-on
point e.g., presence of strong field star contamination  hides it
more strongly than the brightest MS. The other limitation of the
turn-on method is cluster age. As shown by evolutionary calculations,
only for extremely young clusters with ages of order of a few Myr,
the Pre-MS branch is raised sufficiently above the ZAMS and could be
easily recognized in the CMD. For older clusters the Pre-MS branch
deviation from the ZAMS diminishes with increasing age and could be
seen better as a detail of the LF (as it is discussed in
Sec.~\ref{the_sec}). At cluster ages exceeding that of the Pleiades,
the Pre-MS branch could not be identified with confidence both in the
CMD and in the LF and the turn on method cannot be applied
(\citealt{plelf}). 

The other related issue is the fine structure of the stellar LF,
located in the vicinity of the MS turn on point, which can mask the
IMF shape if one does not take it into account. This feature can
produce a false flattening or even depletion in the LF, which are
frequently considered as an evidence of the flattening/turning over
of the IMF (see \citealt{pib} for references). This in turn might
have an important consequence with respect to the IMF universality
and other similar issues. Since luminosity function of the turn on
point depends on cluster age, the position of the LF detail also
varies with time, and may be used as a kind of a standard candle for
age determination from the LF analysis \citep{bep}.

Thus, the MS turn on region of young cluster CMDs is very important
for cluster dating, or should be taken into account when one is
analyzing the LF. It can be easily identified, however, only in the
case of a few selected clusters. The well known examples of such
clusters are NGC~2264, NGC~6530, ONC, and some others. Normally young
and especially remote clusters, which are buried in rich
foreground/background show, neither distinct turn ons nor PMS
branches.

The aim of present study is to reveal  the MS turn on point
information hidden in the existing CCD  observational data of 6 young
open cluster using a new approach, which does not require
observations of blank fields. It should be noted, that  these
clusters were subject of a conventional study, and many of  them
already have age and IMF determination. Unlike these studies, we will
not convert observed LF to the IMF, but in order to avoid the above
mentioned problems, we will construct theoretical LF, which should
properly  reflect the behavior of the observed ones in the vicinity
of the MS turn on  point. Wherever the field star contaminations
could not be determined using the observations of an offset field
region, they are estimated using galactic model for star counts for
the surroundings of the cluster under study. Then  the theoretical LF
will be fitted to the observations by varying the star formation
parameters describing actual stellar population (cluster age, star
formation duration, IMF slope, a percentage of observed field stars).
Considering its success, we hope that this approach can also be
implemented to the other young open clusters having similar or more
deep CCD data, which become presently available (see e.g.
\citealt{CFHT}).

In Section~\ref{the_sec} we detail theoretical approach to LF of a
young  cluster and consider its fine structure in the context of
current study. In  the Sec.~\ref{obs_sec}, we summarize the data used
in the present work.  Sec.~\ref{lf_sec} describes construction of the
observed and theoretical  LFs. We consider here the major effects
which should be taken into  consideration, define the model of
cluster population and its parameters,  and describe the fitting
procedure. In Sec.~\ref{res_sec} we discuss the  derived results, and
summarize them in Sec.~\ref{con_sec}.

\begin{figure*}
 \resizebox{\hsize}{!}
 {\includegraphics[bbllx=55,bblly=350,bburx=515,bbury=635]{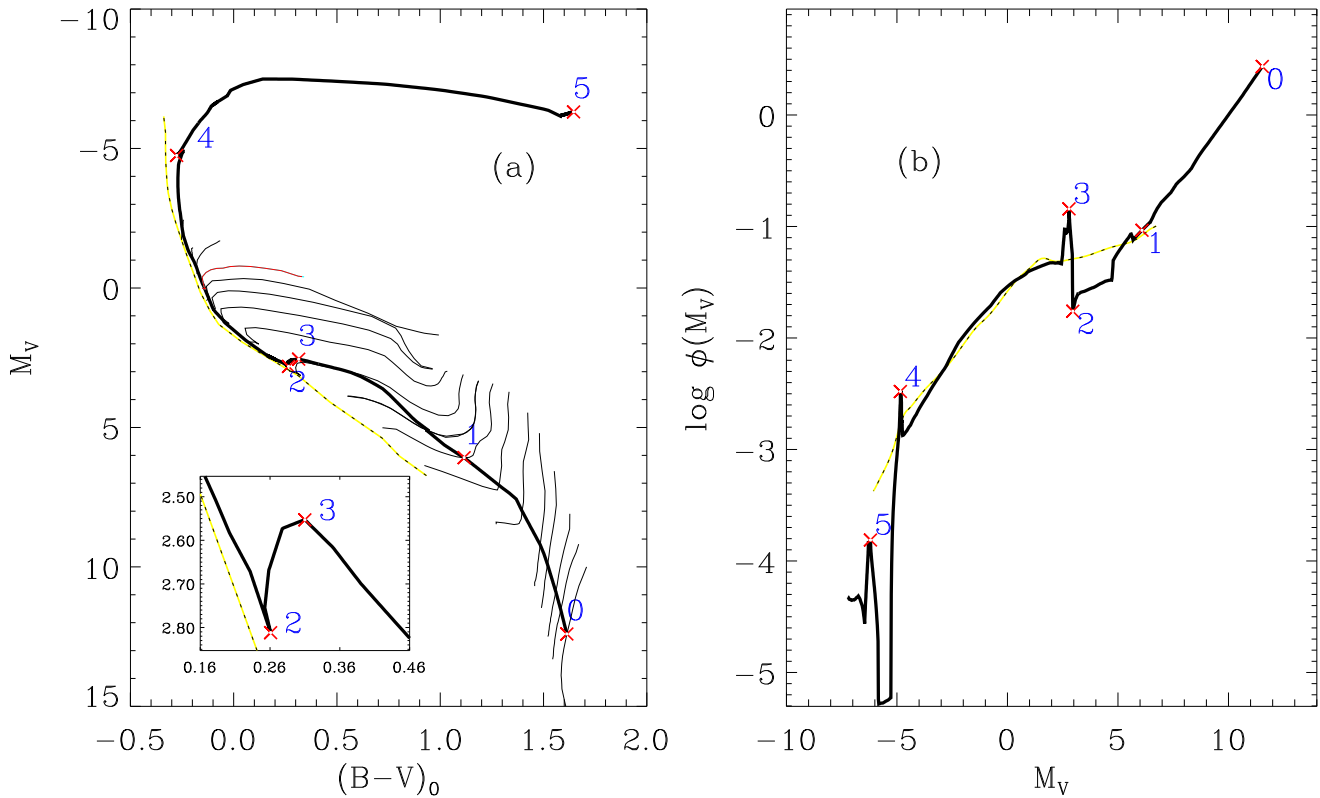}}
  \caption{Theoretical color-magnitude diagram (Panel a) and
  luminosity function (Panel b) of a young open cluster. Solid curves
  are for an isochrone of $\log t =7.0$, including both Pre-MS and
  Post-MS evolutionary stages, and corresponding luminosity function,
  constructed from the isochrone. The dotted curves are the ZAMS and
  "Main Sequence" luminosity function. Thin curves in panel (a) are
  Pre-MS evolutionary tracks of \citet{pas}. The crosses and adjacent
  figures separate different evolutionary stages as it is explained in
  the text. The enlarged MS turn on point area is shown in 
  the panel encapsulated in the CMD.}\label{cmdlf_fig}
\end{figure*}

\section{Theoretical approach}\label{the_sec}

Since stellar mass $m$ can be directly measured in very rare and
specific  cases, the IMF is not observed directly, but is converted
from the observed  distribution of stars over their absolute
magnitudes $dN/dM$, called LF  $\phi(M)$. For a cluster of age $t$
(in years throughout the paper if not  mentioned otherwise), the LF
$\phi_t(M)$ is related to the IMF $f(m)$ via  time-dependent
mass-luminosity relation (MLR) $m_t(M)$ as

\begin{equation} 
\phi_t(M) = \frac{dN}{d\log m}\times\left|\frac{d\log m}{dM}\right|_t 
           =  f(m_t[M])\times\left|\frac{d\log m}{dM}\right|_t.\label{lfdef_eqn} 
\end{equation}     

Since cluster stars evolve off or approach to the MS with a rate
which is  dependent on their mass, apart from their stay on the MS,
the cluster MLR  should also evolve with time, being different at
every moment from the  standard one (usually adopted to be the MLR of
the MS stars). This deviation is especially strong for red giant and
Pre-MS stars. Besides, the "instant"  MLR has a definite fine
structure due to presence of quasi-horizontal  Post-MS and Pre-MS
portions of corresponding isochrone. The structure itself is not too
prominent to influence the LF, but it produces bumps and dips in  the
MLR derivative, and hence in the LF. The strongest fluctuation of
the  derivative occurs near the MS turn off point, but due to the
small number of stars observed there they could not be easily
distinguished from statistical fluctuations of the LF. In contrast,
in the vicinity of MS turn on point, a  sufficient number of stars
coupled with strong MLR derivative bumping  produce a feature, which
can be observed in LFs of star clusters, which are young enough to
display a branch of pre-MS stars. This detail was called by
\cite{pib} as H-feature, as it appears at the location, where the
hydrogen burning starts in the cores of the pre-MS stars.

In order to get an idea on general structure of the LFs of young
star  clusters and their relation to main evolutionary stages of
stars, we show in Fig.\ref{cmdlf_fig} the CMD together with an
isochrone of $\log t=7.0$, and  corresponding LF constructed
according to eq.~(\ref{lfdef_eqn}) with help  of Salpeter IMF ($f(m)
\propto m^{-1.35}$). The isochrone was constructed  from a set of
stellar models described in Sec.~\ref{tlf_sec}. Our  calculations
show, that in general this behavior of the LF does not differ  for
various sets of models, but details (the size and amplitude of the
LF  fragments) may be somewhat different.

As one can see from Fig.~\ref{cmdlf_fig} for young clusters the LF is
non  monotonic even for monotonic IMF, and consists of a number of
monotonic  portions related to different evolutionary stages.
Comparison with  evolutionary tracks shows that segment (0,1) of the
LF corresponds to  convective portion of a Pre-MS track, segment
(1,2) to radiative one, (2,3)  can be identified with MS turn on
point in the CMD, and segment (3,4) belongs to the Main Sequence. Due
to non-monotonic behavior of the isochrone near MS turn on point,
the  segment (2,3) is in fact a superposition of MS and Pre-MS
stellar evolutionary stages and pure MS starts at somewhat brighter
magnitudes. The  last segment (4,5) corresponds to Post-MS stages of
stellar evolution. We  call hereafter for convenience purposes
segment (1,2) as the LF H-dip, and  segment (2,3) as LF H-maximum.
For comparison in Fig.~\ref{cmdlf_fig}, we  show position of ZAMS
models of \citet{tout}, and ZAMS-LF constructed from  these models.
Good agreement between ZAMS- and LFs is observed between
both turn-over points, but beyond MS both LFs differ considerably.

%
%__________________________________________Two column table
\begin{table*}
\begin{minipage}{170mm}
 \caption[]{Specification of studied clusters and used observation data}
 \label{data_tab}
 \begin{tabular}{lccccccclcc}
Cluster &IAU   &$d'$&Area         &Number 
&$V_{lim}$&$V_{cmp}$&$V-M_V$&$E(B-V)$&$\log t$&Refe-\\
        &number&   &($\sq\arcmin$)&of stars&(mag)    &(mag)   &(mag)  &(mag)    &          &rence\\
NGC 2383&C0722--208& 5 & 31.8 &588 &21.8&17.6&13.3 &0.22      &8.45--8.6   & [1] \\
NGC 2384&C0722--209& 5 & 27.8 &256 &20.9&18.9&13.2 &0.22--0.28&7.1--7.3    & [1] \\
NGC 4103&C1204--609& 5 & 19.4 &176 &19.8&17.3&12.5 &0.31      &7.5         & [2] \\
Hogg 15 &C1240--628& 2 &  6.9 &337 &21.7&19.6&16.5 &1.15$^*$  &6.78        & [3] \\
NGC 4755&C1250--600&10 & 38.8 &576 &20.2&18.1&12.9 &0.41$^*$  &7.0         & [4]\\
NGC 7510&C2309+603 & 6 & 38.4 &423 &21.1&16.8&16.0 &1.12$^*$  &$<$7.0      & [5]\\
\end{tabular}
\rule{0mm}{3mm} 
\noindent$^1$\citet{ss99}, $^2$\citet{sc97}, $^3$\citet*{smb},\\
$^4$\citet{sc95}, $^5$\citet{sg91}\\
\rule{0mm}{5mm}
\noindent$^*$~non uniform extinction
\end{minipage}
\end{table*} 

Since the absolute magnitude of a cluster MS turn on point evolves
with time, the H-feature should move faint ward with increasing of
cluster age. The feature not only moves to fainter magnitudes, but
also degrades with time. As calculations of \citet{bep} show it
disappears completely for $\log t \ge 8.2.$ This puts a limit on the
H-calibration applicability. Again the time scale of H-feature
existence depends on specific models, and can be used for Pre-MS
model verification. In principle this makes the LF  of young clusters
a good tool for cluster age determination.

An immediate conclusion, which could be drawn from the above analysis
is that the existence of any bumps/gaps in the LF at or near the MS
turn on point of young clusters should not be regarded as an evidence
of non-monotonous behavior of the IMF. The existence of the Pre-MS
detail in the LF may also result in the mis-interpretation of mass 
spectra of young open clusters. When the limiting magnitude of a
survey  falls within the radiative dip one could observe the LF
turn-over, which  can be wrongly interpreted as a consequence of data
incompleteness or the  IMF turnover. Similarly the steep slope of
convective portion of the LF is  not an evidence of the IMF
steepening.

\section{Observational data}\label{obs_sec}

We have used CCD photometric data of young (age $\le$ several tens
of  Myr) remote and compact clusters presented in papers listed in
Table~\ref{data_tab} (hereafter referred as original papers). The
small  areas occupied by these clusters and their  sufficiently deep
photometry provide high degree of data completeness, which is very
important for LF construction. However, their large distance coupled
with lack of kinematic data prevent  the separation of cluster
members from the field stars and requires  special approach to tackle
this issue.

In Table~\ref{data_tab} the parameters of the clusters under study
are  listed. Cluster designations are shown in  columns~1 and 2, in
column~3 we  list values of cluster angular diameters taken from
\citet{lyn}. The most  of the other parameters are taken from the
original papers mentioned in the last column. Columns~4 through 9
contain area covered by CCD frames, number  of stars within the
frames, limiting and completeness $V$-magnitudes,  apparent distance
moduli and reddening values.  Color excesses, marked  with asterisks
represent average values in variable extinction fields. In column 10
we show log(age) as it was determined in referred studies. We  define
limiting magnitudes of a survey as faintest magnitude of a star in 
cluster sample. Data completeness in magnitude are estimated
following the  procedure described in Sec.~\ref{incompl_sec}.

We compared original data of Table~\ref{data_tab} with \citet{lyn}
catalogue and recent list of reddening values, distance moduli and
ages,  provided by Loktin\footnote{Loktin A.V., Gerasimenko T.P.,
Malysheva L.K.,  2001, Homogeneous Catalogue of Open Cluster
Parameters, Release~2.2. See  also  WEBDA database at URL
http://obswww.unige.ch/webda/}~(private  communication, referred
hereafter as LGM2.2). We find that the extinction  values from
Table~\ref{data_tab} are in excellent agreement with the data  of
both catalogues and the average color excesses from different lists
agree  within a few hundredth of magnitude, except in the case of NGC
7510. In  this case, the extinction data as in LGM2.2 differ from
that of Table~\ref{data_tab} by 0.26 mag. Original cluster ages also
show reasonable agreement (with average spread of the order of a few
tenths in $\log t$) for all clusters except NGC~2383, which according
to both \citet{lyn} and LGM2.2 has an age of a few tens of Myr.

\begin{figure*}
\resizebox{\hsize}{!}
            {\includegraphics[angle=270]{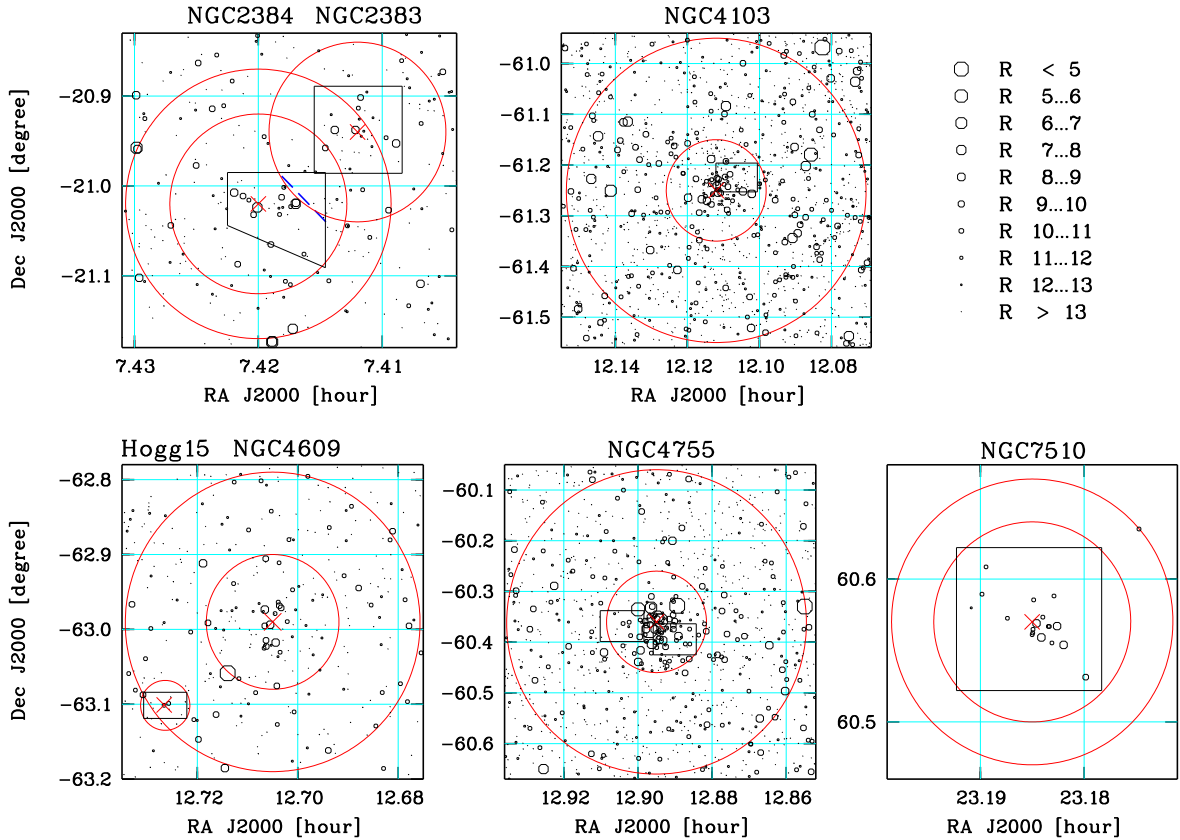}}
\caption{Maps of stellar fields in the vicinity of studied clusters are
constructed with the help of ASCC-2.5, UCAC1, and FONAC catalogues. Size
of the stellar images corresponds to stellar magnitude ($R_{FONAC}$ 
for the case of NGC~7510 and $R_u$ for the others), magnitude 
scale is shown in the upper right corner. Crosses mark cluster centers. 
Large circles denote core and corona areas, solid quadrangles show 
used CCD-frames.}\label{cufmap_fig}
\end{figure*}

Unfortunately such a good agreement is not observed in the case of
cluster  distances. Comparison with \citet{lyn} referred distance
estimations shows  good agreement for all clusters, except NGC~2383
and NGC~2384, where data of Table~\ref{data_tab} are overestimated by
about 1 mag. The new scale of  LGM2.2 indicates even stronger (up to
1.5 mag) disagreement for these and  two more (Hogg~15, and NGC~7510)
clusters. Though these disagreements are unpleasant, it is hardly
avoidable as photometric distances of remote  clusters are derived
from a very steep upper MS. In present study we will  use original
distances, keeping above problems in mind.

The possible uncertainty in cluster age and the MF
slope, which can be introduced by the data inaccuracy can be
estimated from the following considerations. The most important for
our LF-based analysis is accuracy in absolute magnitude $M_V$. It
depends on error in a distance modulus, reddening correction, and on
a correction for metallicity effect. Since no metallicity
measurements are published for clusters under study we can estimate
this correction only indirectly. As the cluster galactocentric
distances span over a range of 7--11 kpc (with the solar value
assumed to be 8.5 kpc), the metallicity radial gradient in the
galactic disk is absent  according to data on Cepheids of
\cite{andr}.  Thus we can accept solar heavy element abundance and
avoid metallicity corrections. The spread of cluster metallicity can
only be be due to galactic disk inhomogeneity, which  according to
\cite{verpis} is about $\Delta \mathrm{[Fe/H]}\approx 0.1$. The
corresponding absolute magnitude uncertainty due to metallicity
effect is of order of $\Delta M_V \approx 0.1$ mag. The standard
accuracy of cluster apparent distance modulus is about 0.2 mag
\citep{ss99}.  The reddening uncertainty according to original papers
is between 0.03 and 0.06 mag and only for NGC~7510 it exceeds
($\sigma_{E(B-V)}=0.09$ mag) the upper limit. These cause the $M_V$
uncertainty to be typically about 0.1 -- 0.2 mag and less than 0.3
mag for the complete sample. Thus we believe that aggregated
uncertainty in absolute magnitude due to metallicity, interstellar
extinction and distance modulus errors does not exceed 0.5 mag. This
corresponds to a random error in $\log t$ of the order of 0.15 both
for turn off and turn on dating methods. According to simulations of
\cite{belt} one can expect an error in the MF slope caused by
$\sigma_{M_V}=0.5$ mag is of order of 0.2.

The original CCD photometry was complemented with photoelectric
magnitudes of brighter stars from the sources referred in original
papers and/or from the WEBDA database. This extends the
range of data completeness from a faint cluster-specific limit,
discussed in Sec.~\ref{incompl_sec} to the brightest in the observed
field stars. Since the data in the $V$-magnitude are the most
complete we use in this study LFs constructed in $V$ only.

As one can see from Table~\ref{data_tab}, stellar fields covered by
the frames are comparable with the cluster sizes provided in
\citet{lyn} catalogue. It should be noted however, that \citet{lyn}
provides only sizes of central parts of open clusters, which 
according to \citet*{kh03} are 2 to 3 times lower than their full
size. Similar conclusions have also been drawn by \citet{nllu} based
on the spatial structure study of the open star clusters. Thus we
should consider our samples only as inner population of the clusters,
enhanced with massive stars \citep{kh03}, and having
flatter luminosity functions and mass spectra than outer population
\citep{dega,degb,degc}.

In Fig.~\ref{cufmap_fig} we show wide neighborhoods of the clusters
under study. They are constructed with the help of all-sky
catalogues  ASCC-2.5\footnote{available at
ftp://cdsarc.u-strasbg.fr/pub/cats/I/280A}  \citep{ascc} for
brightest ($V<13$ mag) stars, and UCAC1\footnote{available at
ftp://cdsarc.u-strasbg.fr/pub/cats/I/268} \citep{ucac}  
[FONAC\footnote{available at
ftp://cdsarc.u-strasbg.fr/pub/cats/I/261}  \citep{fonac} in NGC~7510
case] for fainter stars. In order to give a reader an idea on a
relation of observed frames to cluster geometry we show in the  maps,
positions of cluster centers and borders of cluster cores and
coronae, as determined by \citet{kh03} from data on spatial
distribution of cluster  proper motion members from the ASCC-2.5. For
NGC~2383 \citet{kh03} failed  to detect the core radius and the core
is not shown in the cluster map.  Since brightest stars of Hogg~15
are fainter than limiting magnitude of the  ASCC-2.5, the data on the
cluster center and radius were taken from  \citet*{dias} and
\citet{moffat} respectively.

One can see from Fig.~\ref{cufmap_fig} that regions of two clusters 
(NGC~2383 and Hogg~15) are overlapped with their neighboring clusters. This
should be regarded as a potential source of cluster sample contamination 
and should be kept in mind in further discussions. NGC~4755 is
the only cluster, where observations are carried out in two separated
frames located on both sides of the cluster center. 

\section[]{Observed and theoretical\\* luminosity
functions}\label{lf_sec}

Before comparing the empirical and theoretical LFs, it is necessary
to  correct observed LF for different biases mentioned earlier. Among
the most  serious effects influencing observed LFs are contamination
of a cluster  sample with {\it non-members} and data incompleteness.
Generally the  non-members are field stars projected onto the cluster
area, but in some cases there would be stars from a neighboring
cluster also. The data incompleteness has a strong dependence on the
crowding/density of stars in the field and the stellar magnitude
such that the data is less complete in  crowded regions and towards
fainter magnitudes (see e.g. \citealt{sr91}). All these, as well as
details of procedure for the construction of  theoretical LF and
fitting of theoretical and empirical LFs  will be described below.

\subsection[]{Field star de-contamination with galactic star\\* 
count model}\label{fld_sec}

Depending on cluster distance and on the properties of observed
sample, one  could apply different methods for the selection of
cluster members. For nearby clusters, the kinematic method is the
most common and reliable. For  remote clusters the statistical
selection with the help of adjacent  stellar fields are used. Since
there is no blank fields observed in the  neighborhood of the most
clusters under study, we use the Galactic star count models to
estimate the field star contamination in their directions. Due to
various perturbations (e.g. due to spiral arms) along line of sight,
which can be hardly taken into account in details by the model, this
approach appears to be crude at the first glance, especially for a
study of the fine structure of the LF. A careful analysis indicates
contrary to this apprehension however. It is well known that the
leading role among a number effects influencing the predicted
apparent stellar density, is played by interstellar extinction, as it
is uneven over the galactic disk. For example the patchy behavior of
the extinction is the major reason enabling observation of remote
clusters in transparency windows. This is a basic point of our
approach: since the extinction in a cluster direction could be
derived with much more confidence when compared to an average
extinction value in the disk,  more realistic model predictions can
be made in a cluster field than in an arbitrary stellar field. Below
we describe the model components, discuss the degree of the model
stability against the extinction and stellar density fluctuations,
and compare the model predictions with data on selected test fields.

For each cluster, we compute a theoretical
distribution of field stars with  apparent magnitude $\psi(V)$, which
is specific to the cluster parameters, including galactic
coordinates, interstellar extinction towards the cluster, and
normalized to the area of observed cluster region. We use the  model
described by \citet{ks96}, and \citet{kh97}. The model treats the 
Galaxy as a system, which is symmetric both with respect to rotation
axis,  and to the equatorial plane, and consists of three populations
(thin disk,  the disk hereafter, thick disk and a spheroid). The
population of thin disk  consists of several sub-populations (MS
stars, disk red giants, and super  giants -- stars of I and II
luminosity classes). Each population group is  described by its own
stellar LF and spatial distribution. It is accepted  that population
LF does not depend on position within the Galaxy and is  equal to its
solar neighborhood counterpart. Density distributions of different
populations are represented by \citet{bah} expressions for disk-like
and spheroidal subsystems.

The LF for MS stars brighter than $M_V$ = 13 mag is taken from
\citet{sca}  and \citet{mur}, while for fainter stars ($M_V \leq$19
mag) we used data  given by \citet{jar} for nearby stars. The
fraction of evolved stars, which  have left MS phase (red giants and
super giants) as a function of $M_V$  was taken from \citet{sca}.
These values were modified at $M_V\approx 0.75$ mag, $(B-V)\approx
1.0$ mag according to the data of \citet*{hol} taking into account
red giant clump, an equivalent of Population II horizontal branch for
disk-population core helium burning stars. The LFs of the thick disk
and  spheroid and the star number density normalization \{{\it thin
disk:thick disk:spheroid}\} are \{1:0.02:0.00125\} taken from
\citet{gil}.  Additionally, results by \citet{daco} for globular
clusters were used to  model the influence of the horizontal branch
on the spheroid LF. 

%
%___________________________One column table
\begin{table}
%\begin{minipage}{120mm}
\caption[]{Cluster-specific model parameters}
\label{model_tbl}
\begin{tabular}{lccc}
Cluster &$l^\circ$ &$b^\circ$&$a_V$ \\
        &          &          &(mag/kpc)\\
Hogg 15 &302.0&-0.2&1.00\\ 
NGC 2383&235.3&-2.4&0.38\\
NGC 2384&235.4&-2.4&0.41\\
NGC 4103&297.6&+1.2&0.59\\
NGC 4755&303.2&+2.5&0.92\\
NGC 7510&111.0&+0.1&1.13\\
\end{tabular}
%\end{minipage}
\end{table} 

We adopt the following general parameters of the model. The
galacto-centric distance of the Sun is equal to 8.5 kpc; the ratio
of  axes of the spatial density distribution for the spheroid is
equal to  0.85; the length scale of the disk subsystems and the
height scale  of the thick disk are equal to 4 kpc and 1.3 kpc
respectively. We assume according to \citet{schm} and \citet{sca} 
that the height scale of the disk for MS stars rises from 90 pc to
350 pc  at $M_V$ interval [$2.3, 5.4$] mag, whereas for red giants,
it changes from  250 pc to 400 pc at $M_V=[-0.75, 2.6]$ mag. The
height scale of 90 pc was  used for extremely young population of
super giants.

Young open clusters are located in the galactic disk with irregular
interstellar absorption and their extinction parameters differ from 
average Galactic values. This is a major reason for using in model
calculations a cluster-specific absorption values $a_V$,  which were
derived from \citet{paren} formula
\[
A_V=\frac {a_V \cdot h_Z}{|\sin b|}\left[1 - \exp\left\{\frac {-d \cdot |\sin b|}{h_Z}\right\}\right],
\]
where $d$ is cluster distance from the Sun, $b$ is cluster galactic
latitude, $A_V = 3.1\,E(B-V)$ with $E(B-V)$ taken from
Table~\ref{data_tab}, and $h_Z$ is height scale of the extinction
layer assumed to be equal to  100 pc\footnote{See
paper of \cite{kilma} for recent discussion of the extinction
formula parameters.}. Values of $a_V$ together with cluster galactic
coordinates are shown in Table~\ref{model_tbl}. 

In order to illustrate the decisive role of the 
reddening effect in the model counts construction, we have
confronted  our results to calculations with a standard value of
galactic specific extinction $a^G_V=1.6$ mag/kpc, adopted for the
galactic disk. As shown by the comparison, for all clusters in the
working magnitude range of $V=10-20$ mag, the overestimation of
specific extinction leads to considerable underestimation of the
model density. The corresponding model curves differ in working
magnitude range for different clusters by $\Delta \log N$ = 0.3 --
0.8 (i.e. by 70 to 180 \%). The model stability with respect of
observed errors in the value of specific extinction was evaluated by
variation of values of $a_V$ from Table~\ref{model_tbl} by 0.2
mag/kpc.  According to Sec.~\ref{obs_sec} this value can be regarded
as upper limit of error in specific extinction values in our sample.
We have found that in this case the model counts variations are of
order of $|\Delta\log N|<0.15$ for all clusters under study.

\begin{figure}
\resizebox{\hsize}{!}
{\includegraphics[bbllx=75,bblly=100,bburx=535,bbury=465,
                  angle=270,clip]{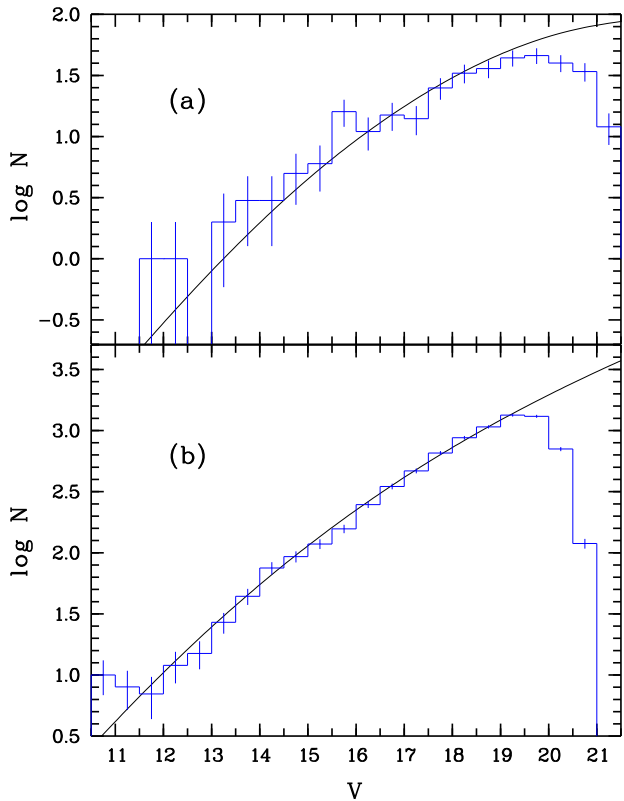}}
\caption{A comparison of stellar counts in test fields,
neighboring to NGC~2383/4 (a), and NGC~7654 (b) clusters, with a
prediction of our model. The histograms show empirical data, the bars
are the Poisson deviations, indicating data uncertainty. The curves
are the model counts.}\label{modcmp_fig}
\end{figure}

In order to estimate the effect of spatial density
fluctuations, we consider galactic spiral arms as the most pronounced
galactic disk sub-structures. We have carried out model calculations
for cluster areas with and without the effects due to spiral arms. We
considered three local arms (Sagittarius-Carina, Orion and Perseus
ones) represented by logarithmic spirals with parameters taken from
\cite{marsuch}. It was assumed, that the density input from the
spiral structure is equal to $\gamma D(R)$, where $D(R)$ is regular
density profile in the disk, and $\gamma$ is a scaling factor, taken
to be equal to 0.1, 0.2, and 0.3. The model calculations show that
for all clusters in the working magnitude range, the influence of the
spiral arms on the stellar density is negligible. The corresponding
increase of $\Delta \log N$  is about 0.05 for $\gamma=0.3$, and less
than 0.02 for $\gamma=0.1$. These values indicate that the effect of
density fluctuation is much weaker than the extinction.

At last we compare our model calculations with
observed stellar counts in selected stellar fields. Unfortunately
among the clusters of our sample only observations of NGC~2383/4 are
supplied with photometry in the blank field containing about 300
stars. The field is about 5\,\arcmin away from the clusters and thus
is partly overlapped with NGC~2383/4 areas (cf.
Fig.~\ref{cufmap_fig}). For model calculations we used data for this
cluster from Table~\ref{model_tbl}. Corresponding distributions are
compared in Fig.~\ref{modcmp_fig}(a). One can observe reasonable
qualitative agreement of empirical and model distributions. According
to $\chi^2$-test the both distributions agree at 80-percent
significance level. Small and generally insignificant access of
brighter stars ($V<16$ mag) can be attributed to the admixture of
NGC~2384 stars.

An excellent opportunity for verifying of our model
provides wide-field CCD-photometry of more than 17\,000 stars in the
field of NGC~7654, published by \cite{pand}. The photometry covers an
area of 40$\times$40 arcmin$^2$, centered at the cluster with a size
of about 24\,\arcmin. This leaves a wide circular blank field area,
enabling not only to confront the model counts and observations, but
also to estimate the stochastic variations of circumcluster fields
themselves. It is worth to note, that NGC~7654 is located only about
1 degree North of NGC~7510, a cluster of present sample.

In Fig.~\ref{modcmp_fig}(b) we show observed and
model distributions for NGC~7654 blank field area. Although
\cite{pand} do not provide explicit value of the cluster radius,
their density profiles show that there is no sufficient cluster
population at $r_0>13.\!\arcmin3$. For construction of the observed
distribution we use all stars located at $r>r_0$, but the stars from
the SE segment of the circumcluster area are not considered. This is
where the neighboring  cluster Czernik~43 having radius of about of
10\,\arcmin is located, according to \cite{KP}. The cluster-specific
model parameters ($d=1380$ pc, and $a_V=1.80$ mag/kpc) were taken
from data for NGC~7654 of \cite{pand}. Since there are indications of
outward increase of the cluster reddening, we used the upper limit
provided by \cite{pand} values. As one can see from
Fig.~\ref{modcmp_fig}(b) the agreement between the model and
observations is good. According to $\chi^2$-test the both
distributions agree at 90-percent significance level in the range
$V=$12--19.5 mag. In order to estimate the degree of stochastic
variations of circumcluster population we arbitrarily divided the
whole blank field into two equal parts and compared them to the
united field separately. The $\chi^2$-test has shown that their
statistics is somewhat different: they are the same as the total
distribution on 85-percent level. This indicates that the model
counts approach to the member selection is at least not worse than
the blank-field method based on an arbitrary sampled test area.

Theoretical differential distribution of stars  with apparent
magnitude $\psi(V)$, derived with help of this model were  used for
construction of reduced LFs as it is described in Sec.~\ref{fit_sec}

\subsection[]{Contamination by an overlapping
cluster}\label{ngb_sec}

The contamination by an overlapping cluster is studied using the
projected  density derived from stellar counts in recent all sky
catalogues. The star counts in the USNO-A2\footnote{Monet D., Bird
A., Canzian B. et al., USNO-A 2.0:A Catalog of Astrometric Standards
(CD-ROM distribution). US Naval Obs., Washington, 1998.} catalogue
is strongly affected by crowding effect in  the areas of Hogg~15 and
NGC~2383/2384 clusters (which is not surprising for Schmidt camera
based surveys in dense fields). To avoid crowding effect we  used
astrograph based catalogue UCAC1. This catalogue is a part of
all-sky  survey UCAC, covers Southern Hemisphere south to
$\delta_{2000} \approx  -20^\circ$, and contains the fields under
study.

NGC~2383/2384 lie at the very edge of the survey UCAC1 area, where
only short exposures provide star counts homogeneity and completeness
at $R_u < 13.7$ mag. We constructed density profiles in a field
centered at the NGC~2384. In order to estimate the proper stellar
density in a cluster, we counted stars in four quadrants separately.
The average observed star densities in the field $\Sigma_f$ and  in
NGC~2384 corona areas $\Sigma_{2384}$ were determined from the counts
in  the first and forth quadrants of the corresponding areas located 
at  $\alpha_{2000} \ge 7\fh42$ and within $0\fdg10 \div 0\fdg15$ and
$0\fdg35  \div 0\fdg60$ from the cluster center which are free from
contamination  of NGC~2383 members. As the values of $\Sigma_f$ and
$\Sigma_{2384}$ do  not strongly vary over the studied area, the
density of NGC~2383 cluster  members can be calculated as
$\sigma_{2383}=\Sigma_{2383}- \Sigma_{2384}$  ($\Sigma_{2383}$ is
counted within NGC~2383 frame). The star density of the  corona of
NGC~2384 itself, then is $\sigma_{2384} = \Sigma_{2384}-\Sigma_f$.
The results are shown in Table~\ref{ngbr_tbl} and one can see that
contamination of NGC~2383 member sample with stars of NGC~2384
cluster computed as  $\sigma_{2384}/\sigma_{2383}$ is about 24\%.

%
%___________________________One column table
\begin{table}
\caption[]{Contributions of field star and neighboring clusters population 
are compared with the projected stellar densities of NGC~2383
and Hogg~15 clusters}
\label{ngbr_tbl}
\begin{tabular}{lrrl}
Population      &\mc{2}{c}{Projected density ($1/\sq^\circ$)}&Designation \\
                &NGC~2383 &Hogg~15                                    &                   \\
                &         &                                           &                   \\
Observed (total)&2724     &6077                                       &$\Sigma_{cluster}$ \\
Field stars     &932      &1870                                       &$\Sigma_f$         \\
Neighbor+field &1273     &2401                                       &$\Sigma_{neighbour}$\\
Neighbor       &341      &531                                       &$\sigma_{neighbour}$\\
Cluster         &1451     &3676                                       &$\sigma_{cluster}$ \\
\end{tabular}
\end{table} 

We cannot estimate the contamination of NGC~2384 member sample with
stars  of NGC~2383 cluster as there are no stars with $R_u < 13.7$
mag within  NGC~2384 frame piece overlapping with NGC~2383 cluster.
Moreover, this  frame piece is small and we can assume insignificant
number of NGC~2383  members there.

Similar procedure was applied to Hogg~15 cluster. The counts were
performed down to $R_u=15$ mag. The average observed density in
corona of NGC~4609, $\Sigma_{4609}$ was computed from the stellar counts at
distance $0\fdg15-0\fdg20$, and average observed density of the
field stars at $0\fdg35-0\fdg60$ from the center of NGC~4609. As
one can see from Table~\ref{ngbr_tbl}, Hogg~15 is contaminated with
NGC~4609 stars by about 14\%. This fraction is smaller than in the
NGC~2383 case, but is not negligible.

\subsection[]{Incompleteness effects}\label{incompl_sec}

It is well known, that completeness of a photometric survey is
achieved only at about 2 mag brighter than $V_{lim}$ (so called
completeness limit $V_{cmp}$). Below $V_{cmp}$ the observed LF does
not reflect behavior of a genuine LF, and this
magnitude range should be excluded from the consideration. It is also
important to be confident, that for clusters, having observations in
two CCD frames both fields have the same completeness limit. 
Unfortunately, no $V_{cmp}$ determinations were made in
original papers. This forced us to determine the $V_{cmp}$ within
this study. Since no test fields were observed in original papers we
are only able to estimate roughly the completeness limit as a position of
apparent maximum in the distribution of stars with $V$ magnitude
within observed frames. Since the incompleteness increases gradually,
the above estimate should be considered as upper limit of $V_{cmp}$.

The estimated values of $V_{cmp}$ are shown in Table~\ref{data_tab}.
For NGC~4755 we find good agreement between $V_{cmp}$ in both
observed frames. Note, however, that the frames do not include the
central area of the cluster, containing the brightest cluster
members. This causes another kind of incompleteness i.e.
incompleteness for bright stars. One should keep this in mind when
the final result for this cluster is discussed. In contrast the frames
of other clusters are located in central regions and no bright star
incompleteness is expected there. 

The other kind of incompleteness, which should be considered is
incompleteness due to overlapping of stellar images in the CCD frame,
which  we call here as "crowding effect incompleteness" (see
\citealt{sr91} for details). It arises due to finite size of stellar
images which depend  on stellar brightness. An image of a star of
some brightness occupies  certain area within the CCD frame, hides
some neighboring stars of lower  brightness, and excludes them from
the statistics, affecting the brightness function $\psi(V)$, a
distribution of stars with apparent magnitude. The  relation between
corrected for the crowding effect brightness function  $\psi_{ce}(V)$
and observed one $\psi_a(V)$ is expressed as
\[
\psi_{ce}(V) = \psi_a(V)\,\left[1 +
\int_{V_{min}}^{V}s(V')\psi_a(V')dV'\right].
\]
Here $V_{min}$ is the magnitude of the  brightest star in the frame, $s(V)$ 
is an area occupied in the frame by an image of a star of magnitude $V$. 
Both $s(V)$ and $\psi_{ce}(V)$ are expressed in units of the frame area.
We have evaluated the crowding effect for each cluster of our
sample. The functions $s(V)$ were found empirically for every frame
from the image statistics as $s(V)=\pi\,r_p^2$, where $r_p$ is
average minimum distance in the plane of a frame from stars of
magnitude $V$ to remaining stars. As we found for the considered
range of magnitudes this effect only negligibly (less than two
percents) changes distribution $\psi_a(V)$. Thus we assume 
$\psi_{ce}(V)=\psi_a(V)$ in our further discussions.

%_______________________________________________Two column table
\begin{table*}
\begin{minipage}{170mm}
\caption[]{Open cluster parameters derived from the fit of the LFs.}
\label{reslt_tab}
\begin{tabular}{lccccccc}
Cluster      &\mc{2}{c}{Fit parameters}&\mc{5}{c}{Derived parameters} \\
             &$M_V$ range &$\chi^2$&$M^H_V$&   $p$      &    $x$        &$\log t$    &$\Delta\log t$\\
             &       (mag)&        &(mag) &            &               &            &       \\
NGC~2383     &$-6.0,\,3.0$&   23.6 & 2.50  &$0.1\pm 0.2$&$2.2\pm 0.4$&$7.1\pm 0.2$&$0.0\pm0.2$\\
NGC~2383$^a$ &$-2.0,\,3.0$&   37.2 & $-$   &$0.2\pm 0.2$&$1.2\pm 0.4$&$8.5$        &$-$\\
NGC~2384     &$-5.0,\,3.0$&   39.6 & 2.00  &$0.3\pm 0.2$&$1.0\pm 0.5$&$7.0\pm 0.2$&$0.2\pm0.4$\\
NGC~2384$^b$ &$-5.0,\,3.0$&   41.5 & 2.00  &$0.2\pm 0.2$&$1.0\pm 0.5$&$7.0\pm 0.2$&$0.2\pm0.4$\\
NGC~4103     &$-3.5,\,2.0$&   24.1 & 1.25  &$0.3\pm 0.2$&$1.5\pm 0.6$&$6.8\pm 0.1$&$0.2\pm0.4$\\
Hogg~15      &$-5.0,\,2.5$&   79.2 & 0.75  &$0.6\pm 0.2$&$1.3\pm 0.4$&$6.7\pm 0.3$&$0.6\pm0.6$\\
Hogg~15$^c$  &$-5.0,\,2.5$&   17.0 & 0.75  &$0.5\pm 0.2$&$1.3\pm 0.4$&$6.6\pm 0.3$&$0.2\pm0.6$\\
NGC~4755     &$-3.0,\,3.0$&   10.8 & 2.50  &$0.2\pm 0.2$&$1.4\pm 0.3$&$7.2\pm 0.2$&$0.2\pm0.2$\\
NGC~7510     &$-6.0,\,1.5$&   13.8 & 0.75  &$0.3\pm 0.2$&$1.2\pm 0.4$&$6.8\pm 0.3$&$0.5\pm0.7$\\
\end{tabular}\\
\rule{0mm}{3mm}
\rule{1mm}{0mm}$^a$ corrected for contamination by NGC~2384 stars,
                    $\log t =8.5$ is accepted from the upper MS isochrone fitting\\
\rule{2mm}{0mm}$^b$ derived in the frame area free from NGC~2383 stars\\
\rule{2mm}{0mm}$^c$ corrected for contamination by NGC~4609 stars
\end{minipage}
\end{table*} 

\subsection[]{Theoretical luminosity functions}\label{tlf_sec}

For construction of a theoretical LF $\phi(M_V)$, we have assumed a
model  of continuous stars formation. It is computed as  \[ 
\phi(M_V) = \int_{t_0}^{t_1} \phi_t(M_V) \lambda(t) \, dt, \]  where
$t_0$ and $t_1$ are minimum and maximum ages of the cluster stars, 
function $\phi_t(M_V)$, computed using equation~(\ref{lfdef_eqn}), is
the  LF of stars with age $t$ ($ t_0 \le t \le t_1$), $\lambda (t)$
is the star  formation rate (SFR) at age $t$. The value of $t_1$,
representing duration since formation of first stars, could be
regarded as the cluster age.

The mass -- absolute magnitude relation $m(M_V,t)$ and its derivative
were  calculated along the isochrone of age $t$ using a cubic spline 
interpolation. For the IMF we considered a power-law representation
$f(m) =  k m^{-x} $, with $k$ to be normalizing factor, and $x$ the
IMF slope to be  determined within this study. We assumed a constant
SFR in our model,  $\lambda(t) = {\rm const}$. The resulting
parameters for each cluster were  drawn from the best fit of the
theoretical and observed LFs.

In order to construct theoretical isochrones and LFs which include
both  Post-MS and Pre-MS stages for ages typical to clusters of our
sample, we  combined Population~I Pre-MS evolutionary tracks of
\citet{dama} for masses  0.1 to 0.8 $m_\odot$, and of \citet{pas} for
masses 0.8 to 6 $m_\odot$, and  Maeder' group Post-MS calculations
\citep{sch92} for $m = 0.8-120 \,  m_\odot$. The grids were properly
tuned to provide a continuous transition  from Pre- to Post-MS ages
and smooth and uniform mass -- luminosity and mass -- radius
relations along the ZAMS. The isochrones were computed from  the
models corresponding to the Population~I chemical abundance ($Y,Z$)
=(0.30,0.02) using linear interpolation.

In order to convert the theoretical isochrones from $\log T_{eff},
\log L/L_{\odot}$ plane to the observed $(B-V)_0, M_V$ plane, we used
bolometric corrections and $(B-V)_0 - \log T_{eff}$ relations from
\citet{schk} tables for the luminosity classes I, III and V.

\subsection[]{Fitting of theoretical and observed LFs}\label{fit_sec}

The following iterative steps are used in the fitting procedure:
\begin{enumerate}
\item {\it Apparent LF construction:} Apparent LF (observed brightness 
   distribution) $\psi_a$ was constructed from the data available for given    frame(s) as smoothed density estimation with a rectangular one-magnitude    wide window, and a step of 0.25 mag. The histogram form of smoothing 
   kernel was selected to apply correction for field star contamination.

\item {\it Construction of a cluster LF:} We assume that apparent LF 
   $\psi_a(V)$ is a composition of a distribution of cluster stars with 
  apparent magnitude $\psi_c(V)$ and field stars brightness function 
  represented by galactic star count model distribution $\psi_f(V)$:
      \[
      \psi_a(V) = \psi_c(V) + p\,\psi_f(V), 
      \]
      where free parameter $p$ denotes percentage of field stars,
      contaminating the sample. With help of apparent distance modulus 
      $V-M_V$ the function $\psi_c$ is transformed to absolute magnitude 
      scale, and is called hereafter as observed  cluster LF $\phi_c(M_V)$.

\item {\it Luminosity function fitting:} Theoretical LF $\phi(M_V)$ 
smoothed in the same way as observed LF was fitted to empirical function 
$\phi_c(M_V)$ within a cluster-specific range of magnitudes. The entropy 
      \[
      \Delta = \sum |\phi_c(M_V^i)\log(\phi_c(M_V^i)/\phi(M_V^i))|
      \]
       was constructed to find the best-fit parameters
      $\{p,x,t_0,t_1\}$. 
\end{enumerate}

Steps (ii) and (iii) were repeated iteratively by varying $p$
until the best agreement (in terms of $\chi^2$ statistics) between
theoretical and observed LFs is achieved. The parameter $p$ was varied 
between 0 and 1 in a step of 0.1.

Internal accuracy of $x,\log t\equiv\log t_1$, and $\Delta\log
t\equiv\log t_1-\log t_0$ was  estimated on the basis of the  kernel
smoothing theory (\citealt{sil}, \citealt{lapko}) for selected
histogram grid parameters (range and step). An accuracy of the field
stars percentage $p$ was estimated from 
\[
\sigma^2_p  =\frac{\sum (p_i-\overline{p})^2/\chi^2_i}{\sum 1/\chi^2_i},
\quad\overline{p}=\frac{\sum p_i/\chi^2_i}{\sum 1/\chi^2_i},
\]
assuming that $\sigma^2_p$ is a second order central moment of a
distribution function equal to that of $1/\chi^2$. 

\section{Discussion }\label{res_sec}
 
The results of the LF fitting are listed in Table~\ref{reslt_tab}.
For each  cluster we show fit parameters: the range of absolute
magnitudes selected for the LFs fit, the best-fit $\chi^2$ value
achieved in the iterations, and derived parameters: position of
H-maximum in theoretical LF $M^H_V$, indicating absolute magnitude of
the turn on point, derived percentage of field star contamination
$p$, the IMF slope $x$, adopted age of a cluster $\log  t$, and age
spread parameter $\Delta\log t$. Note, that parameter errors shown in
Table~\ref{reslt_tab} are in good agreement with the uncertainty
estimations given in Sec.~\ref{obs_sec} on the basis of data accuracy
analysis. The results are discussed  below separately for clusters
under study.

In Fig.~\ref{cmd_fig}, we show the CMDs of clusters under discussion.
The  reddening and distance parameters are taken from
Table~\ref{data_tab} and  fine tuned by the variation of the tabular
values within their accuracy to  reach the best agreement of the
cluster CMD, empirical ZAMS, and  corresponding isochrones. Cluster
proper motion members and non members are  marked, if cluster stars
are bright enough to be included in the ASCC-2.5.  To give an idea
how the adopted distances and reddening values agree with  the
photometry we show in the diagram a position of empirical ZAMS of 
\citet{schk}. In order to illustrate how ages derived from the LF
analyses  confirm with cluster CMDs, we show in Fig.~\ref{cmd_fig},
the isochrones  corresponding to the ages, $t_0$ and $t_1$. 

In Fig.~\ref{lf_fig}, we display the LFs. Observed cluster LF
$\phi_c$,  constructed as described in Sec.~\ref{fit_sec} is shown
with filled  histogram, and corresponding theoretical LF $\phi$
fitted to $\phi_c$ is  shown with a curve (the fitted portion of
$\phi$ is shown with heavy curve, while the rest is marked with thin
curve). For comparison purposes we show  also apparent LFs $\psi_a$
(thin histogram and hatched area), and  contribution from field stars
$\psi_f$ (dotted curve), both displaced by  the value of $V-M_V$.
Below we discuss each cluster as per their appearance in figures
\ref{cmd_fig} and \ref{lf_fig}.

\subsection{Hogg 15}\label{Hogg15_sec}
 
In the cluster CMD, one can indicate the MS turn on point at $M_V^*
\approx  0.5$ mag. Note, that the H-feature should be also found in
the LF in  the vicinity of $M_V^*$, where apparent LF of Hogg~15
stars also shows a  local maximum. The data incompleteness dominates
after $M_V=3$ mag, where  $\psi_a$ gradually decreases in agreement
with Table~\ref{data_tab},  indicating that $M_V^{cmp}=3.1$ mag. The
field star contamination for Hogg~15  is found to be highest amongst
clusters considered here. This effect, as it  is seen from the filled
histogram in Fig~\ref{lf_fig}, keeps the LF practically unchanged at
the brighter portion ($M_V < 0.5$ mag), and hides substantially the
H-feature region.

In order to take into account contamination from an overlapping
cluster NGC~4609 theoretical LF $\phi(M_V)$ was composed of two
populations  belonging to these clusters. Since the clusters reside
at different  distances from the Sun, the stars of NGC~4609 should be
shifted with respect to that of Hogg~15 by the difference of cluster
apparent distance moduli  $\Delta(V-M_V)$. Since $V-M_V$ of the
NGC~4609 according to LGM2.2 is equal  to $11.45$ mag the value of
$\Delta(V-M_V)$ is taken to be equal to $-5.05$ mag. Due to small
fraction of NGC~4609 stars projected on the Hogg~15 area (about 15\%
of Hogg~15 population according to Sec.~\ref{ngb_sec}) the specific
shape of the corresponding mass spectrum  is not important and we
assume that it follows Salpeter law. The age of  NGC~4609 was taken
from \cite{merm81} as $\log t =7.56$. Composite LF is  shown in
Fig.~\ref{lf_fig} with solid curve, while the input from the 
NGC~4609 population is shown with the dot-dashed curve. 

As one can see from Table~\ref{reslt_tab}, taking into account
population of NGC~4609 does not strongly change cluster parameters
(except the age spread), but considerably improves agreement between
theoretical and observed parameters notably reducing $\chi^2$ parameter. 

\begin{figure*} \resizebox{\hsize}{21cm} {\includegraphics{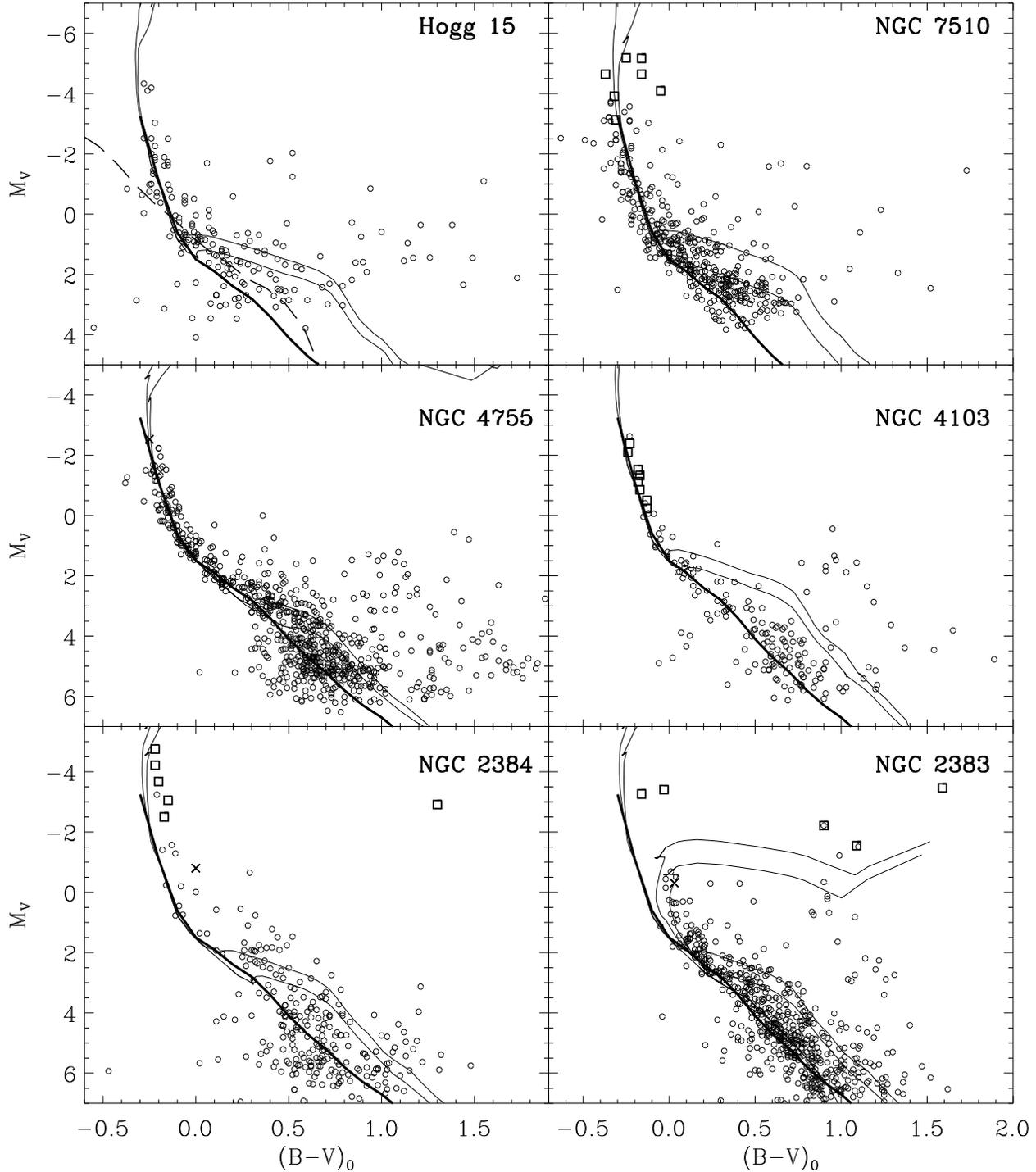}}
\caption{Color-magnitude diagrams of studied clusters. Open circles are 
stars from original observations corrected for adopted $V-M_V$ and $E(B-V)$ 
values. Squares and crosses are proper motion cluster members and field 
stars respectively from ASCC-2.5. Heavy curve is an empirical ZAMS of 
\citet{schk}, and lower and upper thin curves are isochrones corresponding 
to derived age parameters $t_1$ and $t_0$ respectively. For NGC~2383 we show
for comparison both isochrones of $\log t = 8.3$ and 8.5, fitted to its 
upper MS, and isochrones of $\log t = 6.8$ and 7.0, corresponding to age
parameters of NGC~2384, derived from the analysis of its LF. For Hogg~15, 
we show corresponding isochrones as listed in Table~\ref{reslt_tab}. The 
long dash curve is the MS of NGC~4609 as it would appear for Hogg~15 
reddening and distance. }\label{cmd_fig} 
\end{figure*}

Fig.~\ref{cmd_fig} shows that ages $t_0,t_1$ derived from the LF
fitting are in good agreement with cluster CMD. The observed cluster
LF $\phi_c$ also demonstrates a high degree of conformity with
composite theoretical LF. A deep and narrow gap of $\phi_c$ at about 
$M_V\approx 2$ mag shows that the derived input of field stars $p$
is  somewhat overestimated. As shown in Table~\ref{reslt_tab},
Hogg~15 is one  of the two clusters in our sample, where considerable
(although quite  uncertain) age spread of $\Delta\log t = 0.6$ is
detected. This value,  however, is reduced to the insignificant value
of $\Delta\log t = 0.2$,  which is common to the whole sample, if one
takes into account the  contamination from NGC~4609. We therefore
conclude that the age spread in Hogg~15 is introduced mainly due to
admixture of field and overlapping  neighbor' stars, and consider the
cluster parameters as derived for the Hogg~15$^c$ case.

Hogg~15 is the youngest cluster of our sample, with an age of 4 Myr. 
How does this value compare with the earlier published values?  For
the first time, the age of Hogg~15 was estimated by \cite{moffat} as
8 Myr. According to \cite{lyn} the age is 10 Myr.  Recent values come
generally from isochrone fitting of the upper MS: LGM2.2 define it as
$6$ Myr, \cite{smb} provide age estimate of $6\pm2$ Myr, \cite{pc01},
who found initially it as 300 Myr, later reduced this unusually high
value to $20\pm10$ Myr (\citealt{pia02}). Technique of  matching
integrated theoretical and observed spectra of the cluster lead
\cite{ahu00} to the age values $5\pm2$ Myr or 30 Myr. Since MS  turn
off ages are biased toward higher values, our age estimate, based on 
MS turn on point technique, is in fact the lowest one. Note, however,
that  the majority of the above mentioned values agree with the
present one within their accuracy. The IMF slope was determined for
Hogg~15 stars by \cite{smb}, who find it to be equal $1.35\pm0.21$,
which again coincides with the present result.

\subsection{NGC~7510}\label{7510_sec}

It can be seen from the CMD that the regular cluster sequences are
embedded  in a "cloud" of presumably field stars. The astrometric
members also show  definite spread around the MS turn off point. It
may be that some of them  are co-moving field stars. The MS turn on
point can be detected somewhere  between $M_V^* \approx 0.5$ and 1.5
mag. The apparent LF $\psi_a$ shows a  broad plateau between $M_V^*
\approx 0.5$ and 3 mag, with weak maximum at  $M_V = 2.25$ mag. As
indicated by galactic model star counts, it should be  formed mostly
by field stars. The sample demonstrates certain incompleteness below
the maximum, where $\psi_a$ instead of steady increase slowly falls 
down. According to Table~\ref{data_tab} the completeness limit of
original data is $M_V^{cmp}=0.9$ mag. We select $M_V = 1.5$ mag as
faint limit of the LF fitting range based on the value of $\chi^2$
parameter. The agreement between observed cluster LF $\psi_c$ and
theoretical LF is  one of the best amongst the studied clusters.

As one can see from Fig.~\ref{cmd_fig}, the ages $t_0 = 2$ Myr and
$t_1 = 6$ Myr, derived from the LF fitting are in good agreement with
details of the  cluster CMD. The lower pre-MS branch of the cluster
coincides with $t_1$  isochrone. The $t_0$ isochrone fits the
apparent sequence of stars deviating from the MS at $M_V\approx 1$
mag. Stars forming this group do not show  evident concentration to
the cluster center. If, however, this group is not  a random
concentration of field stars, one can regard this as an evidence of
the second star formation event in the cluster.

Excellent agreement between theoretical and observed LFs concerns
only MS.  We are unable to follow the H-feature over its full width,
and are not  certain if the descending portion of the $\psi_c$ is
related to this detail  or it is a consequence of the data
incompleteness of the present survey. Note, however, that cluster age
derived from the assumption that H-maximum  is located at $M_V\approx
1$ mag (just coinciding with the data completeness limit) is in
agreement with the CMD arguments, which is independent of data 
incompleteness. In contrast, the mass spectrum slope, found from
brighter  portion of the LF should be regarded as a confident one.

MS turn off age of NGC~7510 can be found in \cite{lyn} catalogue (10
Myr),  and in LGM2.2 list (38$\pm$2 Myr), \cite{sg91} find the age
$<$ 10 Myr.  Again our age of 6 Myr is the lower estimate of these
values. Mass function  of stars of NGC~7510 was constructed by
\cite{sg98}. It was found, that in  the mass range 1--14 $m_\odot$
the mass function slope $x$ is  $1.1\pm0.2$, which again fits present
results well.

\subsection{NGC~4755}\label{4755_sec}

NGC~4755 is the only cluster of our sample, which is observed in two
frames, located on both sides of the dense cluster center
(Fig.~\ref{cufmap_fig}).  We find that both the frames are similar
with respect to photometric  quality and the completeness issue. So
we consider them together.

\begin{figure*} \resizebox{\hsize}{22cm}{\includegraphics{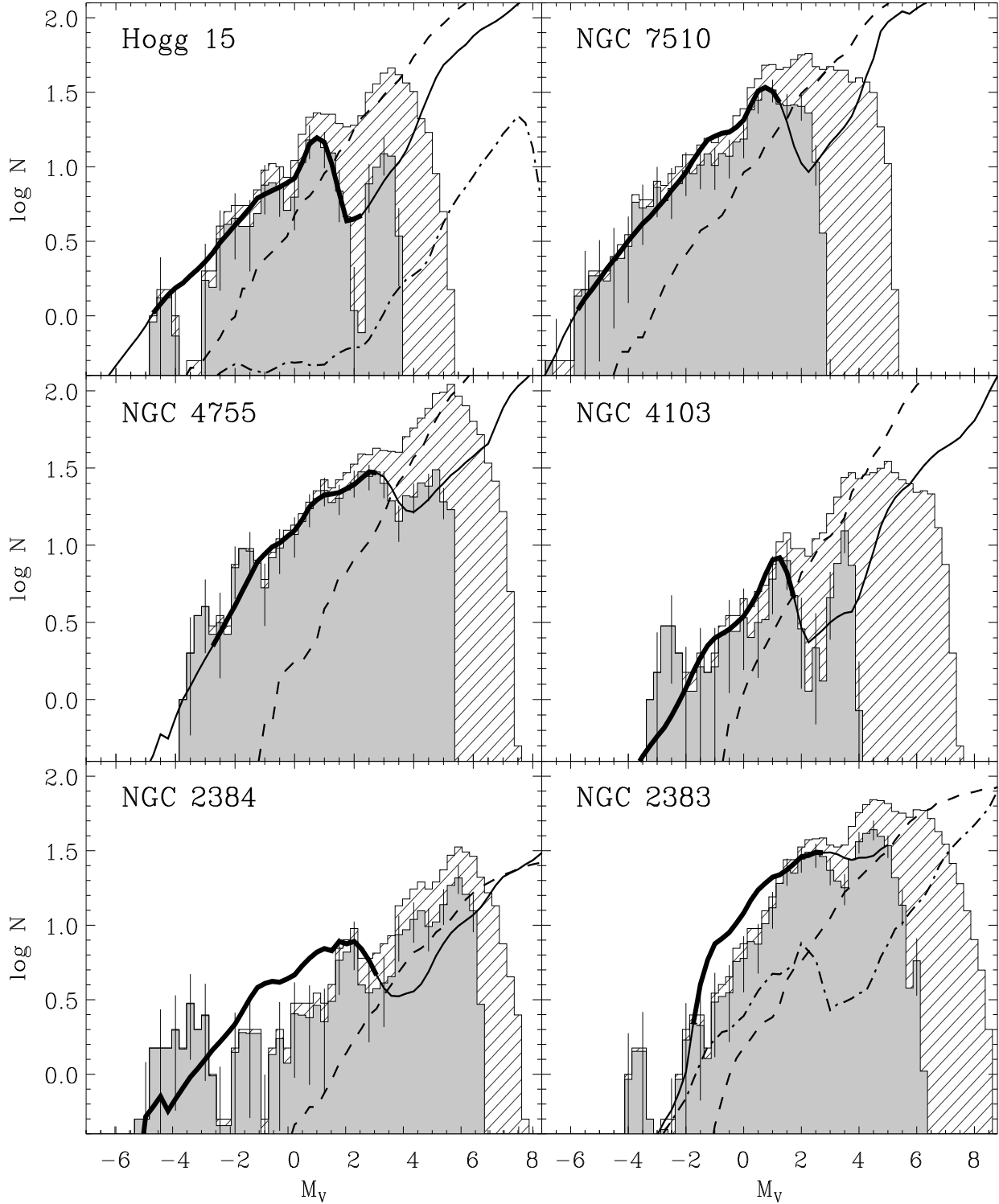}}
\caption{Luminosity functions of the clusters under study. Histograms
show observed LFs: an apparent LF $\psi_a(M_V)$ (hatched), and
cluster stars LF  $\phi_c(M_V)$, corrected for the effect of field
stars (filled), bars are Poisson deviations, indicating data
uncertainties. Curves are  theoretical LFs: cluster star LF
$\phi(M_V)$ fitted to $\phi_c(M_V)$  (solid), and field star
brightness function $\psi(V)$ displaced along the  abscissa by
$V-M_V$ (dashed). Heavy portion of the theoretical LF marks the 
range of fitting. For Hogg~15 and NGC~2383 solid curves show
composite LFs, which take into account the influence of a neighbor,
and dash-dotted curves  show model LF, which mimics neighboring
cluster star population (see  Sec.~\ref{Hogg15_sec} and
\ref{238384_sec} for details).}\label{lf_fig} \end{figure*}

The cluster CMD shows well defined stellar sequences. The MS turn on
point can be seen at $M_V^*\approx 3$ mag, and correspondingly 
H-feature is expected to be located at $M_V = 3$ mag and fainter.
The  uncorrected LF $\psi_a$ shows a step at $M_V=2-3$ mag, related
to  cluster H-feature. The data incompleteness dominates after
$M_V=5.5$ (5.2  according to Table~\ref{data_tab}) mag, where
$\psi_a$ gradually falls down. Correction for field stars for this
cluster can be made with a certain  confidence. By comparing with
model brightness function $\psi_f$, it can be  seen that at brighter
magnitudes the incompleteness can be regarded as  negligible. Thus
selecting $M_V=3$ mag as a faint limit for the LF fitting range, we
are safe from the data incompleteness bias. Since the brightest 
stars, residing in the cluster center and shown in Fig.~\ref{lf_fig},
are  not present in original data (see Fig.~\ref{cmd_fig}), we select
$M_V=-3$  mag as the bright limit of fitting range.  As it is seen
from the filled histogram of Fig.\ref{lf_fig}, we correctly  identify
the H-maximum position, and in spite of the fact that the fitting was
made only for MS portion of the LFs, the theoretical LF reproduces
well the pre-MS H-feature after LF reaches its minimum and turns over
again. However, we are not able to reach the convective portion  of
the LF due to relatively brighter completeness limit of the present
data.

It is seen  that theoretical LF is in good agreement with the the
observed  one $\phi_c$ even outside the fitting range. The derived
age spread of  cluster stars, according to Table~\ref{reslt_tab} is
insignificant, and we  conclude that no evidence of continuous star
formation can be found from  the LF analysis for this cluster. As one
can see from Fig.~\ref{cmd_fig},  the age of 16 Myr as derived from
the LF fitting technique is in good  agreement with the cluster CMD.
The IMF slope coincides within its  accuracy with the Salpeter value.

MS turn off age of NGC~4755 can be found in \cite{lyn} catalogue (7
or 24  Myr), and in LGM2.2 list ($16\pm3$ Myr), \cite{sc95} find that
it is about  10 Myr, and most of pre-MS stars have ages between 3
and  10 Myr. Recent age estimation $10\pm5$ Myr, derived from the
fitting of  upper MS, is given by \cite{san}. For this cluster our
age of 16 Myr is  rather average than the lower estimate of published
values. Mass function  of the NGC~4755 stars was constructed by
\cite{san}. It was found, that in  the mass range 1--13 $m_\odot$ the
slope $x$ is $1.68\pm0.14$, which is significantly steeper than the
value estimated here.

\subsection{NGC~4103}\label{4103_sec}

The observed frame (see Fig.~\ref{cufmap_fig}) covers only a minor
part of  the cluster center. One can therefore expect the results to
be more  uncertain in comparison with the majority of other clusters
under study.

In the cluster CMD, the MS turn on point can be seen at $M_V^*
\approx 1.75$ mag, and the H-feature dip is expected at $M_V = 2-4$
mag, seen as the CMD area of lower density. The apparent LF $\psi_a$
shows week  local maximum at $M_V = 1.5$ mag. The data incompleteness
(see  Table~\ref{data_tab}) dominates at $M_V>4.8$ mag, where
$\psi_a$ gradually  falls down. However, $\psi_a$ flattens after
$M_V=4$ mag, which indicates existence of certain data incompleteness
at brighter magnitudes. To be completely safe against this effect and
taking into account, that the shape of $\psi_a$ at $M_V=2.5-3.5$ mag
is very close to that of model field star brightness function,
indicating high fraction of field stars in  this magnitude range, we
select $M_V = 2$ mag, as a faint limit of the LF  fitting range. The
minimum  of $\chi^2$ is achieved at $p=0.3$, with the  corresponding
parameters shown in Table~\ref{reslt_tab}. 

In spite of the cluster LF $\phi_c$ turning to increase after
$M_V\approx3$  mag, we could not regard the H-feature with same
confidence as in Hogg~15 or NGC~4755, due to poor statistics. It
should be noted, however, that as in  the case of NGC~4755,
theoretical LF, being fitted to the MS observations,  reproduces the
empirical Pre-MS stage reasonably well. Cluster age, derived  from
the LF and MS turn on point analysis shows, that this cluster seems
to  be younger, than it was regarded before, generally from MS turn
off dating  technique (see below). It should be noted, however, that
NGC~4103 is the  case, where the MS turn off technique due to
practically unevolved upper  main sequence of the cluster is able to
provide an upper age estimate only.  In contrast the mass spectrum
slope, found from brighter portion of the LF  should be regarded as
accurate enough. Note some excess of brightest stars ($M_V\approx-3$
mag), which can be however, a consequence of random placement of rare
bright field stars over the cluster region.

All the authors provide  NGC~4103 MS turn off age and are unanimous
in its value. The age is  22--41 Myr according to \cite{lyn},
$24\pm9$ Myr  according to LGM2.2, \cite{sc97} find it to be about 30
Myr, \cite{forb} estimated about 25 Myr and notes, that MS turn on
age should not be much different, and according to \cite{san} the age
is $20\pm5$ Myr. Again we give the lowest value of the cluster age of
about 6 Myr and believe, that both cluster CMD and LF support this
value.  Mass function of stars of NGC~4103 was constructed by
\cite{san}. It was  found that in the mass range 0.7--12 $m_\odot$
the slope $x$ is  $1.46\pm0.22$. Almost the same value is estimated
here. 

\subsection{NGC~2383 and NGC~2384}\label{238384_sec} 

In NGC~2383/2384 pair, the dominating role in the neighbor
contamination is  played by the extended and loose cluster NGC~2384
(see Sects.~\ref{fld_sec}, \ref{ngb_sec}).  In contrast, the central
part of NGC~2384 can be regarded as relatively  free from the NGC
2383 influence. Therefore we start this section with  discussion of
NGC~2384 results.

The CMD of NGC~2384 in its lower part is very fuzzy, while the MS
part of  the diagram is well defined. \citet{ss99} even assume that
stars fainter  than $V=16$ ($M_V=2.8$) mag are mainly field stars, as
can be seen from  the Fig.~\ref{cmd_fig}. Nevertheless, a certain
fraction of cluster members  is observed even at faintest magnitudes.
These are pre-MS stars, forming the upper bound of the observed
sequence at $M_V>2$ mag. In contrast to other  clusters the CMD of
NGC~2384 does not display an extended halo of field  stars, where
cluster MS- and Pre-MS sequences are embedded. Field stars
(presumably giants and late type dwarfs) form a compact clump in the
bottom  of the diagram, with $(B-V)_0\approx0.6$ mag and $M_V\ga4$
mag, leaving  Hertzsprung gap to be unoccupied. The MS turn on point
in the CMD is less  certain than for other clusters. If one
correlates cluster CMD with the  morphology of the observed LF one
should detect it around $M_V^*= 2-2.5$ mag, where cluster
MS-stars certainly cease to appear, and  simultaneously a local
detail, resembling the H-feature appears in the  observed LF. One can
observe, however, that a few of the Pre-MS stars do  appear in the
CMD at brighter magnitudes ($M_V\approx0.5-1.5$ mag).  Their
extension forms an upper bound of a broad Pre-MS branch, extending
down to the limit of observations. The regularity of this feature
implies, that  it is not a random pattern of field stars. They would
be rather  unresolved Pre-MS binaries, or the youngest generation of
NGC~2384 stars.  Note, however, that this brighter position of the MS
turn on point does not  produce anything resembling the H-maximum,
expected to be located near  this point (see Sec.~\ref{the_sec}).

The LF of NGC~2384 demonstrates irregular behavior, and its agreement
with  theoretical LF is worse in comparison to the other clusters
under study. This is confirmed statistically as the $\chi^2$ value is
highest for this cluster (except Hogg~15 case with no correction for
contamination by NGC~4609). One can mention an excess of observed
number of massive stars  ($M_V\approx-4$ mag), and a deficiency of
medium-mass stars ($M_V=-1\div1$  mag) with respect to theoretical LF
$\phi$. Also LF-detail, which we identify with the H-feature is
poorly developed; the observed H-dip is narrower, than the
theoretical one. This behavior is in conformity with
mass segregation scenario in the case if the observed frame covers
the central part of the cluster only. The derived slope of the mass
function is similar to values found by \cite{degb} for central parts
of rich LMC clusters NGC~1805 and NGC~1818. The LF irregularity could
be attributed then to incomplete member statistics. The higher extent
of the cluster comparing to the observed frame naturally explains
other peculiarities of NGC~2383/4 pair like presence of early type
stars in the NGC~2383 field and existence of underdeveloped H-feature
in its LF (see below), or weak excess of bright stars in the
neighboring blank field (see. Sec.~\ref{fld_sec}).

The stability of NGC~2384 parameters against contaminating
effect  of NGC~2383 was checked using data from the part of NGC~2384
frame, located  outside NGC~2383 area (see Fig.~\ref{cufmap_fig}). As
one can see from  Table~\ref{reslt_tab}, there is no difference
between derived parameters of  NGC~2384 in both cases. Thus we can
regard the influence of NGC~2383 member contamination on the derived
parameters of NGC~2384 to be negligible.

Color magnitude diagram of NGC~2383 is more populated than that of
the neighboring NGC~2384. There is a well defined evolved portion of
the MS compatible with cluster age $\log t=8.5$, which is in
agreement with an  estimate of $\log t=8.45-8.6$ by \citet{ss99}
and differs from the ages  $\log t=7.4$ and 7.17 derived by
\citet{lyn} and  LGM2.2 respectively.  This contradiction arises due
to the presence of bright ASCC-2.5  kinematic members with ages of a
few tens of Myr. The MS turn on point,  corresponding to this younger
age can not be easily identified in the CMD. Nevertheless the LF
indicates presence of the H-detail even in the  apparent LF $\psi_a$.
The MS-band at fainter magnitudes ($M_V\ga 2$ mag) is  widened and
unlike the other clusters studied here, it is not due to  field
giants extending the MS from the side of fainter magnitudes, but it
is due to bright stars forming the high luminosity envelope of the
MS. 

In the corrected LF $\phi_c$, the H-feature becomes more pronounced,
leaving the H-dip however to be somewhat underdeveloped (more shallow
and narrow  than in the corresponding theoretical LF). The cluster
age derived from the  LF fitting corresponds to lower age estimation
coming from the brightest MS  stars. The value of $\chi^2$ indicates
that the LF fit is more reliable,  than in the case of NGC~2384. Note
the unusually steep value of the IMF  slope $x$ = 2.2, estimated for
this cluster. In spite of formally satisfactory results of the LF
fitting there are several points, indicating that NGC~2383 carries
signs of a dual nature, which were not taken into consideration. For
example, evolved upper  MS indicates that NGC~2383 is about 300 Myr
old (see Fig.~\ref{cmd_fig}),  while LF indicates its age close to
that of NGC~2384. The relatively young  age of NGC~2383 is also
supported by the presence of bright MS kinematic  members. Possibly
this is the main reason for assigning younger age to the  cluster by
\cite{lyn} and LGM2.2. This confusion can be removed with the help of
a model of an overlapping neighbor, with the only difference from 
the case of Hogg~15, that in this case degree of contamination with
NGC~2384 stars is higher (see Sec.~\ref{ngb_sec}). This is also
supported by  independent data of the ASCC-2.5, which indicates, that
these clusters are  located close not only in the sky, but also have
similar proper motions (see \citealt{kh03}). So a probable proper
motion member of one cluster can  also well belong to the other, and
the brightest kinematic members of NGC~2383 could be actually 
projected members of NGC~2384. This can also be supported by the fact
that  numerous MS-stars of NGC~2383 forming upper envelope at $M_V\ga
2$ mag,  could be actually Pre-MS stars of NGC~2384. Apart from
coincidence with positions of corresponding isochrones in the CMD
this idea is also supported by spatial distribution of these stars.
They tend to be located  in NGC~2383 frame within circle of its
neighbor. The number of these stars is compatible with the degree of
contamination derived from stellar counts and they might be
responsible for steeper IMF slope found for NGC~2383, when neglecting
influence of NGC~2384.

In order to take into account contamination from the overlapping
cluster  theoretical LF of NGC~2383, $\phi(M_V)$ was composed of two
populations  belonging to both clusters. Unlike to the case of
Hogg~15, NGC~2383  and NGC~2384 reside at the same distance within
the errors  ($\Delta(V-M_V) =0.1-0.3$ mag). So no correction is
required for the LF magnitude scale while constructing the composed
theoretical LF. We assume,  that the mass spectrum of NGC~2384 stars
does not vary with location and  the case of NGC~2384$^b$ in
Table~\ref{reslt_tab} can be considered as its  representative. The
corresponding values of $t_0$ and $t_1$ were also taken  from the
Table~\ref{reslt_tab}. Since no developed H-feature exists in the LF
at $t\sim$300 Myr we were forced to use the LF of NGC~2383 in a
customary, way i.e. for determination of $x$ and $p$ parameters.
Cluster age was found from the MS turn off point and fine tuned to 
satisfy the $\chi^2$-test.  The lowest $\chi^2$-value was achieved at
$\log t=8.3$, which is somewhat less than isochronic age of  $\log
t=8.5$ (Fig.~\ref{cmd_fig}), determined from the Post-MS isochrone 
fitting. Due to above mentioned reasons we regard this disagreement
as insignificant and show in the Table~\ref{reslt_tab} the turn-off
age. The best-fit composite LF for the case NGC~2383$^a$ is shown in
Fig.~\ref{lf_fig} with solid curve, while the contaminating model LF
of the NGC~2384 is shown with the dash-dotted curve.

As one can see from the Table~\ref{reslt_tab}, the proposed model
provides  less accurate fit of theoretical LF to observed LF than the
previous one.  On the other hand, it harmonises the full observation
scope (star counts,  cluster CMDs and LFs). As a result, with the age
determined from the MS turn off point calibration, the IMF slope has
been reduced to the value close to  the Salpeter's one with the value
of $p$ almost unchanged.

In the light of above discussions, it is not surprising that the
cluster parameters derived in the present study are in good agreement
with those derived in the original paper by \cite{ss99}. For example,
our values of $x$ practically coincide with those of \cite{ss99}, who
found $x=1.3\pm0.15$  for NGC~2383 and $x=1.0\pm0.15$ for NGC~2384.
Similarly, in agreement with \cite{ss99}, we regard that NGC~2383
ages listed by \cite{lyn} as 25--41 Myr or by LGM2.2 as 16 Myr are
underestimated, while the NGC~2384 ages  (1--10 Myr according to
\citealt{lyn} and $8\pm1$ Myr according to  LGM2.2) are in fair
agreement with our result. However, again the age of  NGC~2384
derived from LF-fitting is lower than that derived by \cite{ss99} 
from the upper part of the MS. 

\cite{ss99} conclude that in spite of their spatial proximity
clusters NGC~2383/2384 do not constitute a physical pair in the sense
of common  origin. Present study not only supports this point but
makes it more  stronger. In spite of the fact that both clusters do
overlap in the plane  of sky, reside approximately at the same
distance from the Sun, and have  similar proper motions, they are
very different in all other respects e.g.  cluster geometry,
morphology, environment, stellar contents, and age and  can not be
regarded as twins.

\section{Conclusions}\label{con_sec}

The major aim of the present study was to elaborate tools, which
provide comprehensive investigation of remote young open star
clusters having accurate CCD-photometry. The effort was focused on
revealing of the Main Sequence turn-on areas, important both from the
point of young cluster dating and from the luminosity and mass
function analysis. To enable the statistical selection of cluster
members for vast set of open cluster CCD observations with no data on
adjacent blank fields, and thus, to involve them to luminosity and
mass functions studies we proposed a new approach to field star
removing technique based on using of galactic disk star count model.
As we found, due to high confidence in interstellar extinction
determination, open clusters are especially suitable objects for this
technique, providing selection results at least of the same quality
level as the standard method of the blank field does. The proposed
approach could be especially valuable for deep observations of young
clusters, planned to be observed in the frame of present-day surveys
(see e.g. \citealt{CFHT}), which are able not only to reach turn-on
regions of selected young clusters, but also reveal complete
H-details down to convective portions of luminosity functions.

The main conclusion from the  present study of the 6 such clusters
residing at helio-centric distances  of 2--4 kpc is that their 
CCD observations coupled with the Galaxy model star counts, wide area
statistics provided by all sky catalogues, and theoretical LFs fitted
to the observations are sufficient to study their population,
construction of the detailed luminosity and mass  functions, and age,
provided the photometry is deep enough to reach the LF H-feature. The
direct comparison of observed and theoretical LFs instead of 
converting them to stellar mass spectra is of principle importance in
this  approach. It provides a standard candle, which can be  used for
a reliable cluster dating. This is very difficult otherwise, since 
the upper MS of young clusters usually implemented for this aim, are
too steep, and degree of stellar evolution is as a rule insignificant
to provide reliable cluster ages. In fact it just provides an upper
estimates for the cluster age. The main conclusions can be written
as:

\begin{itemize}
\item Contamination of cluster members with field stars is the most 
  important factor influencing the lower part of cluster CMD and LF.
  For the clusters under study, field star contamination varies in
  the range of 20--50\%.

\item The overlapping clusters may considerably distort the observed
  LFs.  The "neighbor" contamination, in addition to field stars, is
  equal to 14\% in the case of Hogg~15 and to 24\% in the case of
  NGC~2383.

\item The LF H-feature was found well beyond statistical noise in all
  the clusters under study. For most of the clusters its location
  agrees well with the theoretical prediction. We believe, that in
  the case of NGC~2384, where H-feature displays underdeveloped
  structure, this can be explained by complexity of the area, and
  spatial incompleteness of the data. The false H-feature observed in
  NGC~2383, which certainly is too old to demonstrate it, is due to
  contamination of the cluster field with NGC~2384 stars.

\item Cluster ages derived in the present study are as a rule several
 times lower than those determined from the fitting of the
 theoretical isochrones to the turn off parts of the MS. We believe,
 that our ages are more accurate, than those derived from the upper
 MS. Note that although we used continuous star formation model, the
 derived durations of star formation events are only insignificantly
 differ from zero.

\item Stellar mass spectra of studied clusters are well represented
 with a power law with slopes, which as a rule agree  with Salpeter'
 value, within errors. One should keep in mind however, that since
 the observed data concerns the central parts of clusters, the
 derived mass spectra might be different from the IMF due to mass
 segregation effect. The unusually flat mass function of NGC~2384 can
 be treated e.g. as an indirect evidence of higher extent of this
 cluster and of presence of mass segregation effect.

\end{itemize}

\section*{Acknowledgments}

We thank the anonymous referee for useful comments. This work was
partly supported by the RFBR grant No 01-02-16306. One of us (AEP) is
thankful to the Indian National Science Academy, New  Delhi for
funding the visit to India in the fall of 2000, when this work was
initiated. ANB acknowledges the financial support of the INTAS 
(grant INTAS YSF 00-152).

\bsp
\label{lastpage}
\end{document}